\documentclass[a4paper,english,journal=nalefd]{achemso}
\usepackage[T1]{fontenc}
\usepackage[latin9]{inputenc}
\usepackage{array}
\usepackage{amsmath}
\usepackage{units}
\usepackage{multirow}
\usepackage{amstext}
\usepackage{graphicx}
\usepackage{setspace}

\makeatletter

\title{Charge doping induced phase transitions in hydrogenated and fluorinated graphene}
\author{Tim Wehling}
\affiliation{Institute for Theoretical Physics, Universität Bremen, Otto-Hahn-Allee
1, 28359 Bremen, Germany}
\alsoaffiliation{Bremen Center for Computational Material Science, Universität Bremen,
Am Fallturm 1, 28359 Bremen, Germany}
\email{wehling@itp.uni-bremen.de}
\author{Bernhard Grundkötter-Stock}
\author{B\'alint Aradi}
\affiliation{Bremen Center for Computational Material Science, Universität Bremen,
Am Fallturm 1, 28359 Bremen, Germany}
\author{Thomas Niehaus}
\affiliation{Department of Theoretical Physics, University of Regensburg, 93040 Regensburg,
Germany}
\author{Thomas Frauenheim}
\affiliation{Bremen Center for Computational Material Science, Universität Bremen,
Am Fallturm 1, 28359 Bremen, Germany}

\usepackage{babel}
\begin{document}
\begin{abstract}

  We show that charge doping can induce transitions between three distinct
  adsorbate phases in hydrogenated and fluorinated graphene. By combining ab
  initio, approximate density functional theory and tight binding
  calculations we identify a transition from islands of C$_8$H$_2$ and
  C$_8$F$_2$ to random adsorbate distributions around a doping level of $\pm
  0.05$~e/C-atom. Furthermore, in situations with random adsorbate coverage,
  charge doping is shown to trigger an ordering transition where the sublattice
  symmetry is spontaneously broken when the doping level exceeds the
  adsorbate concentration. Rehybridization and lattice distortion energies make
  graphene which is covalently functionalized from one side only most
  susceptible to these two kinds of phase transitions. The energy gains associated with the clustering and ordering transitions exceed room temperature thermal energies.

\end{abstract}
\maketitle

Low dimensional materials provide unique opportunities to manipulate their
properties by chemical means. Graphene in particular is a zero band gap Dirac
material which can be turned into a wide band gap insulator by
hydrogenation\cite{Elias2009} or
fluorination\cite{Robinson2010,nair2010fluorographene}. Partially functionalized
graphene offers a unique chance to tune optical and electronic transport
properties between disordered Dirac material and insulating characteristics by
varying the adsorbate concentration\cite{Elias2009} and the real space
arrangement of the adsorbates \cite{nair2010fluorographene,Leenaerts_PRB10,Eriksson_PRB10,Abanin2010,Trauzettel_PRL11,Ciraci_PRB2011}. Similarly, electron correlation phenomena
including magnetism \cite{Grigorieva_NPhys2012} and
superconductivity\cite{savini2010} can be expected to be most sensitive to
adsorption patterns in chemically functionalized graphene. It is thus crucial to
be able to tune real space arrangements of adsorbates for on-demand
functionalization of graphene.

\begin{figure}%
\includegraphics[width=0.5\linewidth]{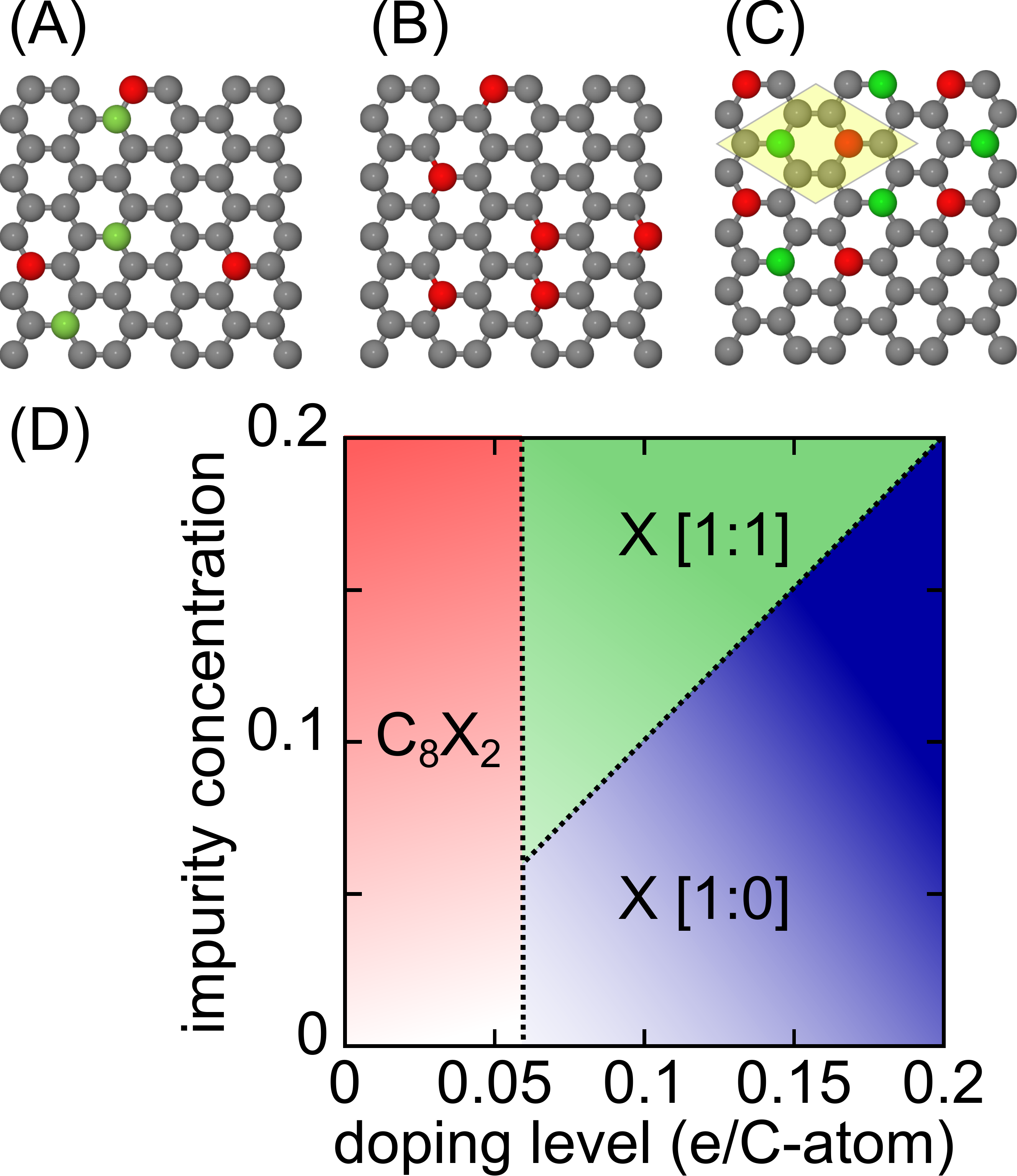}

\caption{Top views of hydrogen and fluorine adsorption patterns on
  graphene. Adatoms are colored according to their sublattice position (red --
  on sublattice $A$, green -- on sublattice $B$). (A) X[1:1] structure. Fully
  random coverage with equal population of both sublattices.  (B) X[1:0]
  structure. Hydrogen and fluorine atoms binding randomly to one sublattice
  only. (C) Phase separation into C$_8$X$_2$ islands and pristine graphene. The
  shaded area marks one C$_8$X$_2$ unit. (D) Schematic illustration of phase
  diagram of H and F adatoms adsorbed to graphene.}%
  \label{fig:structs_phase_diag}%
\end{figure}

Interestingly, field theoretical studies suggested various structural phase transitions in dilute graphene adsorbate systems including instabilities towards Kekul\'e and sublattice symmetry broken patterns \cite{Shytov2009,Cheianov_PRB09,Cheianov2009,Cheianov2010,Abanin2010}. It remained, however, unclear which of these transitions could be realized experimentally, particularly in situations with sizable adsorbate coverage ($\sim 5\%$ to $20\%$). In this letter, we show that adsorption patterns of hydrogen and fluorine atoms on graphene can be largely manipulated by charge doping. By combining ab initio
density functional theory (DFT), the density functional tight binding scheme
(DFTB) and tight binding calculations we find that charge doping can induce
transitions between phases with homogeneous adsorbate distribution over the
entire sample and separation into clean graphene and areas with maximum
adsorbate coverage (Fig. \ref{fig:structs_phase_diag}). We furthermore find
that in case of homogeneous adsorbate distribution, charge doping can trigger an
ordering transition where the sublattice symmetry is spontaneously broken.


In general, the interplay of several mechanisms determines the stability of
graphene derivatives: First, covalent adsorbates like H or F lift their C
bonding partners out of the graphene plane and rehybrizdize them from $sp^2$ to
$sp^3$. There are furthermore electronic energies associated with bond formation
as well as electronically mediated interactions between
adsorbates\cite{Shytov2009,Cheianov_PRB09,Cheianov2009,Cheianov2010,Abanin2010,HuangPRB12,Solenov_PRL13}. We
show that rehybridization and lattice distortion energies make graphene which is
covalently functionalized from one side only most susceptible to the above
mentioned phase transitions.


To study the influence of electron and hole doping on adsorption patterns of
hydrogen and fluorine on graphene, we have investigated their stability by
quantum mechanical simulations. The dependence of adsorption energies $E_{\rm
  ads}$ on adsorption patterns and charge doping has been calculated according
to
\begin{equation}
E_{\textrm{ads}}=\frac{E_{\textrm{G:X}}-E_{\textrm{G}}}{n_{\textrm{X}}}-\frac{1}{2}E_{\textrm{X}_{2}}.\label{eq:XoG_IntE}
\end{equation}
Here, $E_{\textrm{G:X}}$ is the energy of the doped graphene sheet
with the adsorbed atoms X (X=H or F), $E_{\textrm{G}}$ is the energy of the
doped graphene sheet of the same size without the adsorbates, $n_{\textrm{X}}$
is the number of adatoms, and $E_{\textrm{X}_{2}}$ is the energy of the adatom
dimer. For all random adsorbate distributions considered below, each
$E_{\textrm{ads}}$ presents an average over 20 configurations.

The quantum mechanical calculations for obtaining the total energies were
carried out using the DFTB+ program package\cite{AradiDFTB2007} (version 1.2.2)
with the parametrization sets mio-1-1\cite{mio-1-1} for H-adsorption and
pbc-0-3\cite{pbc-0-3} for F-adsorption. The doping has been simulated by
employing the virtual crystal approach (VCA) \cite{Freysoldt2009,Makov1995}. The
various adsorption configurations have been relaxed until the forces on the
atoms were smaller than $10^{-4}$~Hartree/Bohr.  In order to check the
reliability of the results, selected configurations have been recalculated using
ab initio all electron DFT calculations as implemented in the FHI-AIMS
code\cite{blum_ab_2009} (version 081912) using the provided default tight basis
sets for the atoms and the PBE exchange correlation functional. Here, the structures were relaxed using similar force criteria as used in DFTB+. For the k-point sampling we have used 4x4x1 and 2x2x1 Monkhorst-Pack meshes in the DFTB+ and FHI-AIMS
calculations, respectively.  

As shown in the supplementary material, the variations of
the adsorption energies between different patters as obtained from DFTB are very close to
the DFT results for hydrogen adsorption. In case of fluorine adsorption DFTB
generally overestimates the penalty for building sublattice polarized
configurations with respect to sublattice symmetric configurations by
approx.~$0.2$~eV/atom. However, this does not change our statements about the doping
dependent phase transition between the sublattice polarized and sublattice
unpolarized adsorption patterns qualitatively. These transitions should indeed
occur at lower doping concentrations than predicted by the DFTB results. The
absolute values of the adsorption energies are energetically too much favorable
for H-adsorption and not favorable enough for F-adsorption with DFTB with
respect to our ab initio DFT reference calculations. However, energy differences
between all patterns as well as trends with doping are reasonably described by
our DFTB simulations. (See the supplementary material for a detailed comparison
of the DFTB and DFT results.)



\begin{figure}%

 \includegraphics[width=0.90\textwidth]{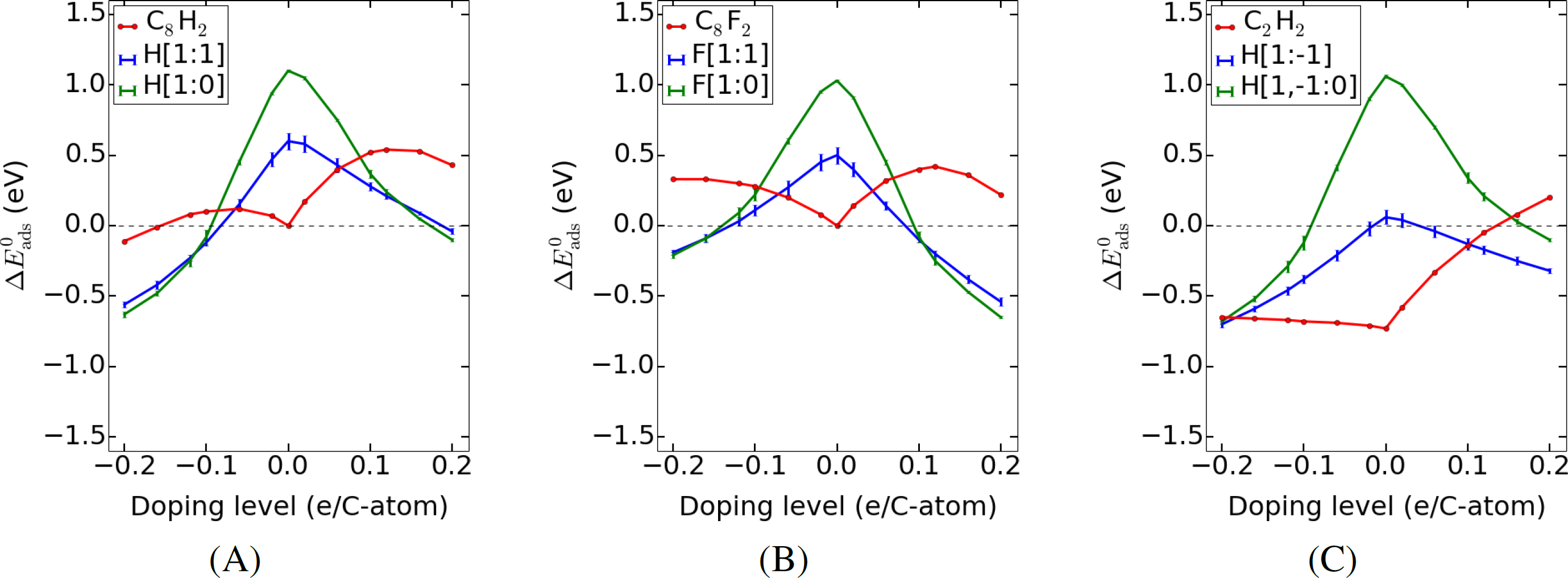}
  \caption{Average relative binding energies (per adatom) for hydrogen (A,C) and
    fluorine adatoms (B) on graphene as function of charge doping for different
    adsorbate arrangements. (A,B) Single sided functionalization. (C) double
    sided hydrogenation. Energies are given with respect to the adsorption
    energy of C$_8$H$_2$ (A,C) and C$_8$F$_2$ (B) at zero doping. For the
   structure with random adsorbate distribution, each data point in (A,B,C)
    refers to an average over 20 different adsorption configurations with 20
     hydrogen/fluorine atoms on a (10x10) graphene cell (200 carbon atoms). Error
     bars give the standard deviations. \label{fig:HFaG-eDopedGrapheneAverage}}
 \end{figure}

First, we consider the adsorption of H and F on one side of graphene. Previous
DFT calculations showed that the C$_8$X$_2$-structures (Fig. \ref{fig:structs_phase_diag}C shaded area) correspond to the upper concentration limit for H/F single
side adsorption on graphene \cite{Haberer2011,Robinson2010}. We thus compare
$E_{\textrm{ads}}$ for the ordered C$_8$X$_2$ structures to graphene with $10\%$
adsorbed hydrogen and fluorine adatoms\footnote{$100\%$ adatom coverage refers
  to one adatom per carbon atom.} in fully random (X[1:1],
Fig. \ref{fig:structs_phase_diag}A) and fully sublattice polarized but otherwise
random (X[1:0], Fig. \ref{fig:structs_phase_diag}B) adsorption patterns in
Fig. \ref{fig:HFaG-eDopedGrapheneAverage}. At zero doping the C$_8$X$_2$ structures
are by several 100 meV per atom more favorable than the X[1:1] or X[1:0]
structures. This finding is in line with the tendency of H and F to aggregate
when adsorbed on graphene \cite{Besenbacher2006,Haberer2011,Robinson2010}.

The situation changes, however, drastically with charge doping. With increased
doping level the adsorption energy decreases for the quasi random X[1:1] and
X[1:0] patterns but not for the C$_8$X$_2$-structures. Therefore, carbon-adatom
bonds are strengthened in the X[1:1] and X[1:0] patterns according to the sign
convention of Eq.~(\ref{eq:XoG_IntE}). At doping levels above $\pm 0.05$
e/C-atom the X[1:1] configurations become more favorable than C$_8$X$_2$ with
adsorption energy differences exceeding several 100 meV/X-atom. For H and F
coverage on the order of $10\%$, charge doping can thus destabilize the
separation into clean graphene and C$_8$X$_2$ islands. Doping therefore induces
a phase transition from separated C$_8$X$_2$ islands to other adsorption
patterns like the fully random X[1:1] pattern with energy gains largely
exceeding room temperature.

We now turn to situations of doping beyond $\pm 0.05$ e/C-atom. With increasing
charge doping, the adsorption energy difference between sublattice polarized
X[1:0] and unpolarized X[1:1] patterns decreases and eventually even reverts
sign. The fully sublattice polarized adsorption patterns become lowest in energy
at electron and hole doping above $\sim 0.1$~e/C-atom for $10\%$ hydrogen and
fluorine coverage, respectively
(Fig.~\ref{fig:HFaG-eDopedGrapheneAverage}A,B). This tendency towards sublattice ordering corresponds to the phase transition suggested in Ref. \cite{Abanin2010}. Notably, for one-sided
adsorption at strong charge doping, we find an energy gain of 60-100 meV/X-atom
upon sublattice
ordering. 
These binding energy differences clearly exceed room temperature thermal
energies and suggest that a second doping induced phase transition between
X[1:1] and X[1:0] structures should be, therefore, achievable even in room temperature
experiments.

Clearly, the electron-hole asymmetry in the doping dependence of adsorption
energies (\ref{fig:HFaG-eDopedGrapheneAverage}A and B) differs between randomly
hydrogenated and fluorinated graphene, which reflects the difference in the
polarity of the C-X bond.

We furthermore considered hydrogenation from both sides. Here, full
hydrogenation of graphene, i.e.~$100\%$ hydrogen coverage with H atoms binding
to sublattices A and B above and beneath the graphene sheet, respectively, is
possible and leads to the formation of graphane \cite{Elias2009}. We consider
two sided random hydrogen distributions at $10\%$ coverage (i.e.~$5\%$ above and
below), where H is either sticking to sublattice A only (H[1$\bar{1}$:0]) or
where it binds to sublattice A from above and B from below (H[1:$\bar{1}$]). The
adsorption energies follow a qualitatively similar trend with charge doping as
in the case of single side functionalization
(Fig. \ref{fig:HFaG-eDopedGrapheneAverage}C). With increasing charge doping, hydrogen
adsorption energies in the H[1$\bar{1}$:0] and the H[1:$\bar{1}$] configurations
become more negative, while there is an increase in $E_{\rm ads}$ for graphane
with charge doping. However, the formation of graphane is more favorable than
random H[1$\bar{1}$:0] or H[1:$\bar{1}$] adsorption patterns over a much wider
doping range than in the case of single side hydrogenation. Moreover, there is
no doping induced transition towards a sublattice ordered H[1$\bar{1}$:0] state
in the doping range under investigation. The H[1:$\bar{1}$] structure remains
more favorable than the H[1$\bar{1}$:0] state even up to doping levels of $\pm
0.2$ e/C-atom. Therefore, doping induced adatom phase transitions are much
easier realized in single side covalently functionalized graphene.


We now aim to identify the microscopic mechanisms behind the charge doping
dependent emergence of different adsorbate patterns found above. There are two
distinct contributions which determine the dependence of binding energies on
doping and adatom patterns: first, adsorbate interactions mediated by the
band structure energy of the graphene $\pi$-electron
system\cite{Shytov2009,Cheianov_PRB09,Cheianov2009,Cheianov2010,Abanin2010,HuangPRB12,Solenov_PRL13}
and second strain and rehybridization energies. Only the latter contributions
distinguish between adatom adsorption from one versus two sides. In the fully
sublattice polarized patterns the energy difference between one and two-sided
adsorption is almost an order of magnitude smaller than for patterns with equal
sublattice population (c.f.~\ref{fig:HFaG-eDopedGrapheneAverage}A and
C). The rehybridization and strain contributions are thus larger in situations,
where both sublattices are covered. Previous DFT calculations on graphene with
two hydrogen adatoms have shown that binding of two hydrogen atoms to two
neighboring C atoms on different sides of the graphene sheet is by 0.5~eV more
favorable than binding on the same side \cite{Boukhvalov2008}. For H pairs on
second or third nearest neighbor positions the energy differences between single
and double side adsorption are at least a factor of two smaller. Thus, the large
strain and rehybridization energy differences in patterns with coverage of
sublattices A and B originate from pairs (or also larger clusters) of hydrogen
atoms, which bind to nearest-neighbor carbon atoms.

\begin{figure}%
\includegraphics[width=1.0\columnwidth]{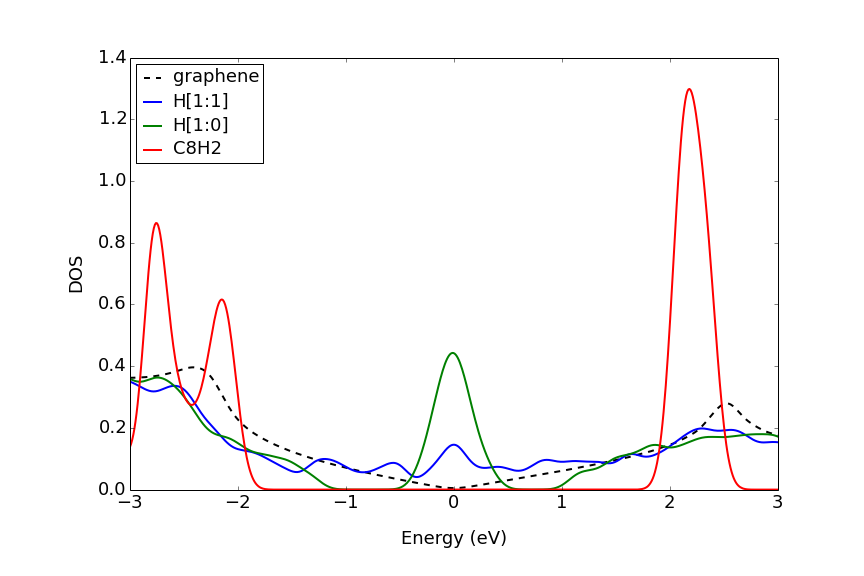}
\caption{Density of states per unit cell of graphene with different levels of
  hydrogenation: pristine graphene, $10\%$ hydrogenation with fully random
  H[1:1] and sublattice ordered H[1:0] configuration, and C$_8$H$_2$. The Fermi
  level of each configuration has been shifted to zero.}%
\label{fig:dos}%
\end{figure}

The band structure energy contribution to the adsorption energy differences can be estimated from the electronic density of states shown for the C$_8$H$_2$, H[1:1], and H[1:0] configurations in Fig. \ref{fig:dos}. In contrast to the other patterns, the spectrum of C$_8$H$_2$ exhibits a gap which makes its formation favorable in the undoped case. Due to the gap, charge doping of C$_8$H$_2$ is however associated with larger band structure energies than for the other two adsorbate configurations. Thus, C$_8$H$_2$ clusters are destabilized and H[1:1] or H[1:0] patterns become more favorable at a certain doping level (c.f. Fig. \ref{fig:HFaG-eDopedGrapheneAverage}). The electronic DOS of the H[1:1] structure is gapless around the Fermi level, while the H[1:0] structure exhibits are large peak in the DOS at $E_F$. While the large DOS at $E_F$ in the undoped state of the H[1:0] pattern makes this pattern unstable against structural (as well as possibly magnetic) reconstructions, doping of the H[1:0] pattern requires only small amounts of band structure energy. Therefore, the H[1:0] pattern becomes eventually most favorable at large doping.

More quantitatively, band structure energy differences can be described in terms
of a tight-binding (TB) model $H=H_{\rm gr}+H_{\rm imp}$. Here, $H_{\rm gr}=-t\sum_{i,j}c_i^\dagger c_j$ is the nearest-neighbor tight-binding
Hamiltonian of graphene, where $c_i^\dagger$ ($c_i$) creates (annihilates) an
electron at site $i$ and $t=2.6$\,eV is the nearest-neighbor hopping
parameter. The adsorbates are taken into account through the Hamiltonian $H_{\rm
  imp}=\epsilon_{d}\sum_{i'}d_{i'}^{\dagger}d_{i'}+V\sum_{i'}\left(
  d_{i'}^{\dagger}c_{i'}+\mathrm{h.c}\right)$, where $d_{i'}^{\dagger}$
($d_{i'})$ creates (annihilates) an electron at a defect level at a defect site
with adsorbant ($i'$). The sum runs over all defect sites. The parameters $V=6$\,eV, $\epsilon _{d}=0$ for hydrogen and $V=6$\,eV, $\epsilon _{d}=-2$\,eV for
fluorine have been fitted to our DFTB results and are in line with previous DFT
calculations \cite{Wehling2010}.  The TB simulations were performed on
supercells containing 1800 C-atoms and are averaged over 100 impurity
configurations each.

\begin{figure}%
\includegraphics[width=\columnwidth]{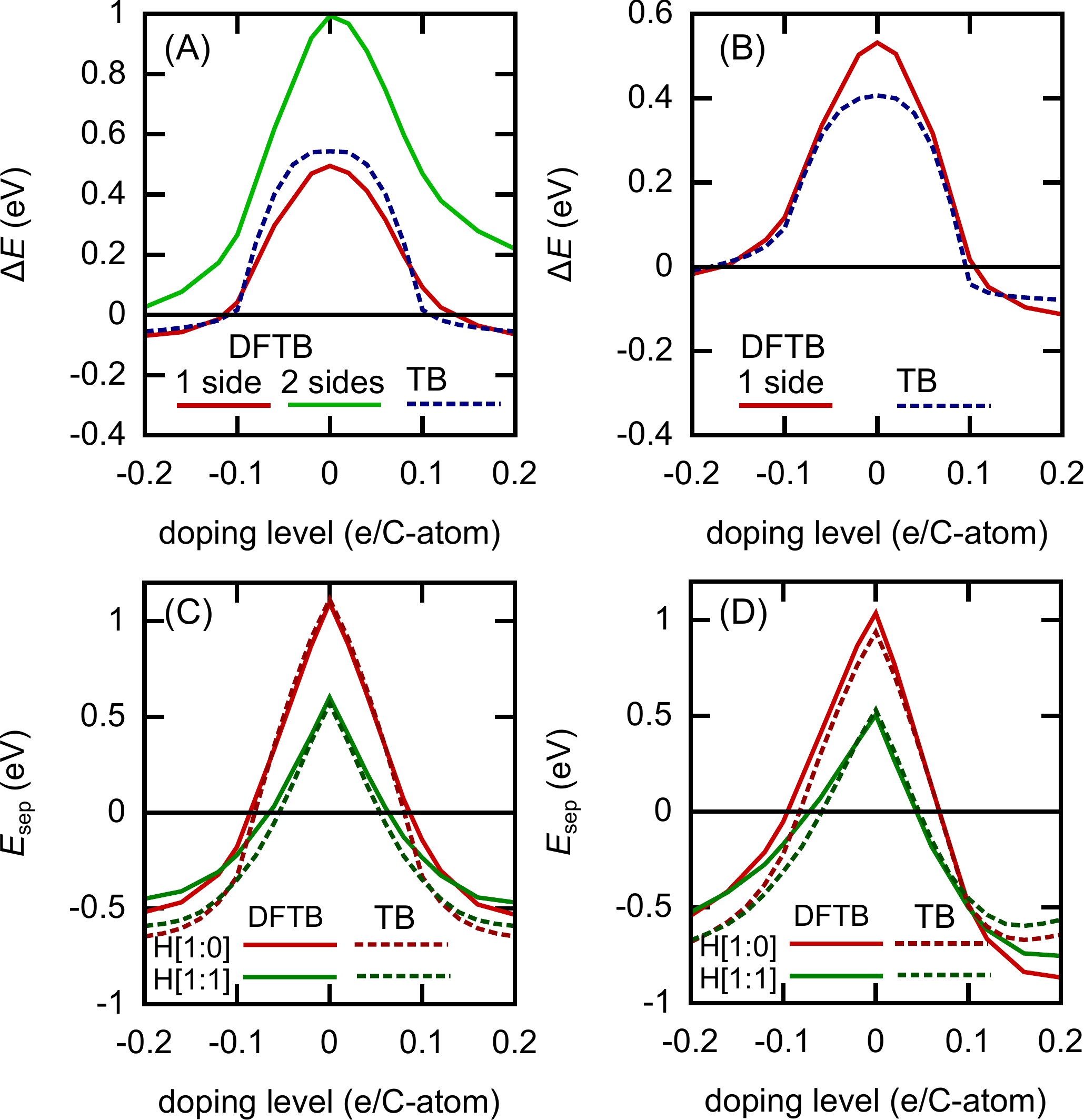}
\caption{Adsorption energy differences $\Delta E$ (Eq.~\ref{eq:XoG_DeltaE})
  between fully random X[1:1] and sublattice polarized X[1:0] configurations of
  H (A) and F adatoms on graphene (B). Solid lines are the DFTB results; TB
  dashed. In (A), the adsorption energy difference of hydrogen between
  sublattice unpolarized (H[1:$\bar{1}$]) and polarized (H[1$\bar{1}$:0]) patterns
  is also shown for two side adsorption. (C,D) Energy gain $E_{\rm sep}$ upon phase separation
  from X[1:0] or X[1:1] structures to C$_8$X$_2$ islands. The
  sample averaged H/F coverage is $10\%$ in all cases.}%
\label{fig:DFTB_TB_asymm_En}%
\end{figure}

The energy difference
\begin{eqnarray}
  \Delta E & = & E_{\textrm{ads}}\left(\text{X[1:0]}\right)-E_{\textrm{ads}}\left(\text{X[1:1]}\right)\label{eq:XoG_DeltaE}
\end{eqnarray} 
between sublattice polarized X[1:0] and unpolarized X[1:1] patterns as obtained
from the DFTB and TB model are compared in Fig.~\ref{fig:DFTB_TB_asymm_En} A,
B. While the TB results significantly deviate from the DFTB results for double
sided hydrogenation, they reproduce the DFTB adsorption energy differences for,
both, single side hydrogenation and fluorination, very well. The energy
difference $\Delta E$ decreases with charge doping of any sign and changes sign
at similar doping levels in DFTB and TB.

We furthermore evaluated the energy difference associated with the phase separation into graphene and
C$_8$X$_2$ islands
\begin{equation}
  E_{\rm sep}=E_{\rm G:X}-\left[(1-4\, c)\,E_{\rm G}+4\,c\,n_{\rm C}\,E_{\rm C8X2}\right]
\label{eq:Esep}
\end{equation}
for hydrogenated and fluorinated graphene at $c=10\%$ coverage and various
doping levels
(Fig.~\ref{fig:DFTB_TB_asymm_En}C,D). 
The number of carbon atoms in the graphene sheets is indicated by $n_{\rm C}$. There are clearly quantitative differences between $E_{\rm sep}$ as obtained from DFTB and the TB model. Nevertheless, the TB models reproduce the doping levels, where the phase separation into
graphene and C$_8$X$_2$ becomes less favorable than the X[1:1] or X[1:0]
configurations, at least qualitatively correct. Also away from this phase transition,
energy differences are at least captured qualitatively correct by the TB model.



The TB model is therefore used to extrapolate the DFTB results and to construct the
charge doping and impurity concentration dependent phase diagrams of H and F
adsorbed to graphene. To this end, we calculated total energies of graphene at
several concentrations $c$ of adsorbed H and F, several charge doping levels
$n$ and evaluated the phase separation energies $E_{\rm sep}$ and energy gains
upon sublattice ordering $\Delta E$ according to
Equations (\ref{eq:Esep}) and (\ref{eq:XoG_DeltaE}). Thereby, we consider the electron doped case only, since the TB model of H on graphene is particle-hole symmetric and the tendency towards phase transitions in fluorinated graphene is strongest on the electron doped side.

\begin{figure}%
  \includegraphics[width=1.0\columnwidth]{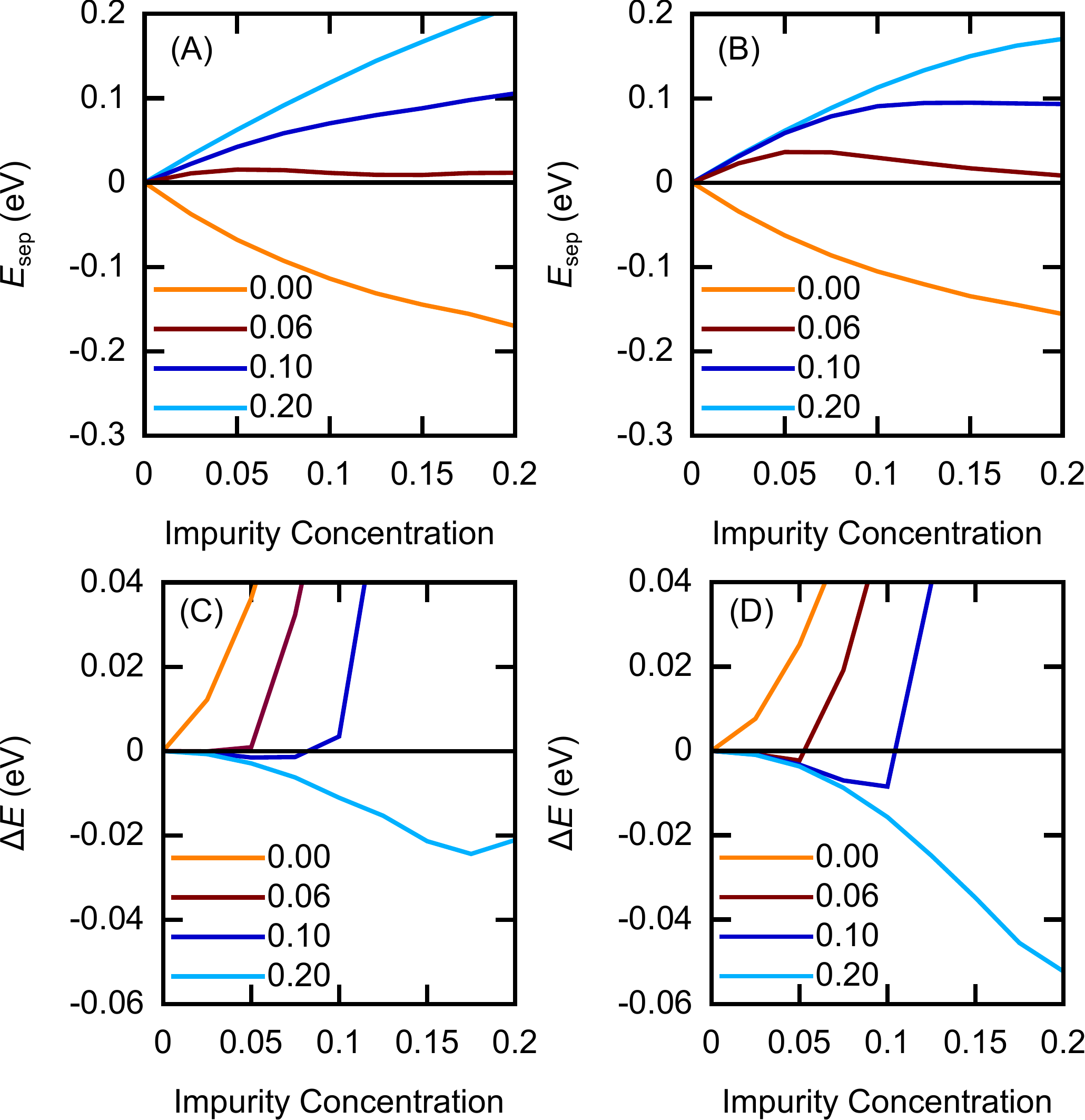}
  \caption{(A,B) Energy gains upon phase separation, $E_{\rm sep}$ from fully
    random X[1:1] coverage to C$_8$X$_2$ islands for single hydrogenated (A)
    and fluorinated graphene (B).  (C,D) Energy gains upon sublattice symmetry
    breaking (X[1:1]$\to$X[1:0]) for hydrogenated (C) and fluorinated (D)
    graphene. Different curves refer to different charge doping levels between 0
    and $0.2$e/C-atom. Energy gains per unit cell are given. }%
  \label{fig:phase_sep}%
\end{figure}

As can be seen from Fig. \ref{fig:phase_sep}A, the TB model suggests that the
graphene-C$_8$H$_2$ phase separation occurs at sufficiently small charge doping
($n< 0.06$~e/C-atom) for all adatom concentrations ($c<20\%$) considered,
here. The energies associated with the phase separation can exceed room
temperature thermal energies by more than an order of magnitude. The situation
reverts around $n\sim 0.06$~e/C-atom. Here, the randomly hydrogen covered
configurations appear more favorable by energies per unit cell which can again
exceed room temperature already at adatom concentrations $c<5\%$. Once the
doping exceeds the adatom concentration ($n>c$), the sublattice symmetry broken
H[1:0] structures become more favorable than the fully random H[1:1]
configurations. For fluorine (\ref{fig:phase_sep}B) a qualitatively similar
picture emerges. However, the tendency towards the destruction of the C$_8$F$_2$
islands and the sublattice ordering is stronger here.

Taken together, our DFTB and TB calculations suggest the phase diagram shown in
Fig. \ref{fig:structs_phase_diag}D. There is a phase separation into graphene and
C$_8$X$_2$ at sufficiently small charge doping. For charge doping exceeding
$n\sim 0.06$~e/C-atom the adatoms distribute over the entire sample, where
sublattice symmetry breaking becomes favorable when the doping level exceeds the
adatom concentration ($n>c$). In the vicinity of both transition lines more
complex phase separated adsorption patterns might emerge. This can be seen from
the concave shape of $E_{\rm sep}$ at $n\sim 0.06$ e/C-atom
(Fig. \ref{fig:phase_sep}A,B) as well as the steep increase in $\Delta E$ as
soon as $c>n$ (Fig. \ref{fig:phase_sep}C,D).

Hydrogen and fluorine adsorption on graphene are highly sensitive to external
charge doping. Under which experimental circumstances could switching between
different adsorption patterns be expected? Electrostatic doping
\cite{Mannhart2003,Morpurgo2011} allows to achieve carrier concentrations on the
order of $10^{14}$~e/cm$^2=1$~e/nm$^2\approx 0.03$~e/C-atom. According to our
results, this alone is not enough to break the tendency towards
graphene-C$_8$X$_2$ phase separation. Chemical doping, for instance by means of
alkali, earthalkali, or rare earth intercalation between graphene and its
substrate, however, allows for electron doping up to $\approx 0.1$ e/C-atom
\cite{Dresselhaus_AdvPhys_2002,Rotenberg_Seyller_APRES2010,Michely_NJP2012}. Thus,
intercalated graphene samples are the most promising systems to explore the rich
variation of covalently functionalized graphene systems with charge doping. In
these electron doped systems particularly fluorine adatoms are highly
susceptible to doping induced phase transitions.

\subsection{Acknowledgements}
We thank the European Graphene Flagship for financial support as well as
HLRN (project hbc00011) for computer time. T.O.W. would to thank Bj{\"o}rn Trauzettel and Vladimir Fal'ko for useful discussions as well as KITP Santa Barabara for hospitality, where ideas presented in this work were conceived.

\subsection{Supplementary material}

As explained in the main article, the adsorption energies for the various
hydrogen and fluorine adsorption patterns have been calculated using the total
energies obtained by calculations with the DFTB+ program package
\cite{AradiDFTB2007}. For every doping concentration and adsorption type (X[1:1]
and X[1:0]) we considered 20 randomly generated configurations, which have been
relaxed with the force criterion described in the main article.  In order to
test the reliability of the data, ab initio calculations for a few selected
adsorption pattern had been carried out using the FHI-AIMS
code\cite{blum_ab_2009}. 

\begin{figure}
  \parbox[b]{0.48\columnwidth}{
    \centering{%
      \includegraphics[width=0.48\columnwidth]{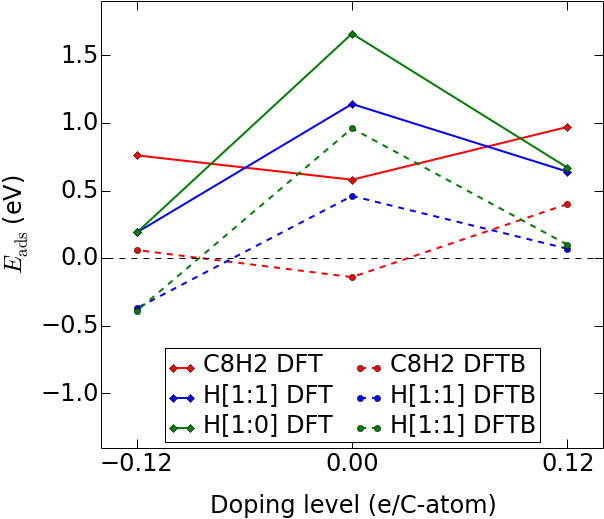}
      {\small (A)}}%
  }
  \hfill
  \parbox[b]{0.48\columnwidth}{
    \centering{%
      \includegraphics[width=0.48\columnwidth]{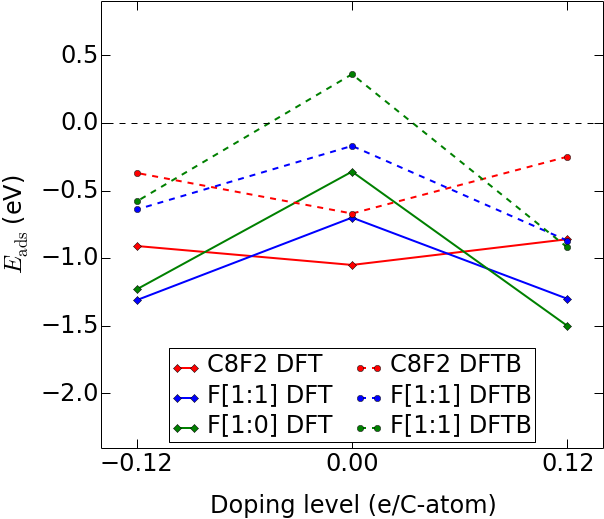}
      {\small (B)}}%
  }
  \caption{Adsorption energies for selected (A) hydrogen and (B) fluorine
    adsorption patterns as calculated by ab initio DFT and by the DFTB methods.}
  \label{fig:aims}
\end{figure}

\begin{figure}
  \parbox[b]{0.48\columnwidth}{
    \centering{%
      \includegraphics[width=0.48\columnwidth]{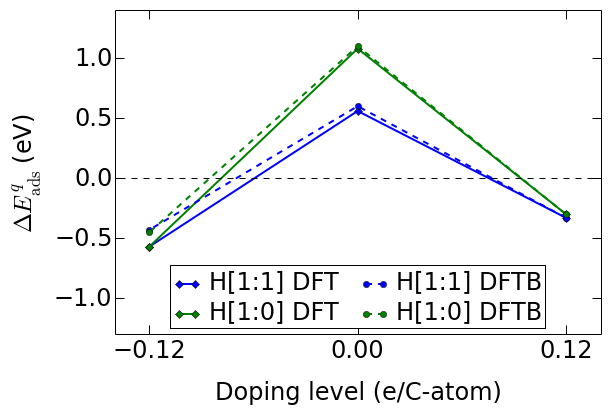}
      {\small (A)}}%
  }
  \hfill
  \parbox[b]{0.48\columnwidth}{
    \centering{%
      \includegraphics[width=0.48\columnwidth]{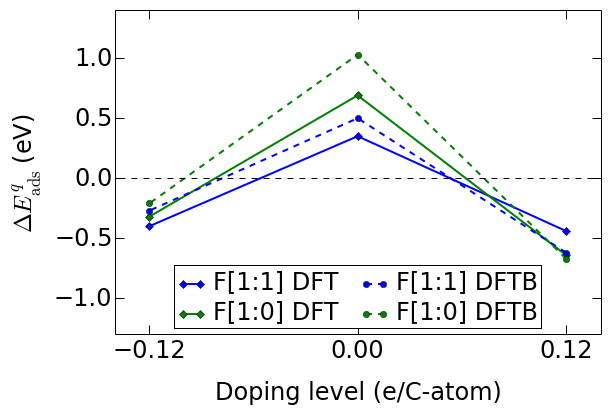}
      {\small (B)}}%
  }
  \centering
  \caption{Adsorption energies relative to the adsorption energy of the (A)
    C$_8$H$_2$ and (B) C$_8$F$_2$ configurations at various doping levels for
    selected configurations as calculated by the DFTB method and ab initio DFT
    method.}
  \label{fig:dftb-aims}
\end{figure}

As shown in Fig. \ref{fig:aims}, the absolute
adsorption energies differ significantly from the DFTB values. Ab initio DFT
calculations predict the hydrogen adsorption being unfavorable in the entire
investigated doping range, while fluorine adsorption is favorable for all
investigated doping levels. It is important to note, that the adsorption
energies are calculated with respect of pristine graphene and isolated H$_2$ or
F$_2$ molecules, which does not resemble the experimental conditions for
hydrogenation and fluorination. In contrast to the absolute energies, the
relative energies of the various configurations are very similar in both, DFTB
and DFT. Both methods predict the C$_8$X$_2$ configuration being less favorable
at doping $\sim \pm0.1$~e/C-atom. Also, both predict that the sublattice
polarized configurations become more favorable than the sublattice symmetric
ones beyond these doping concentrations. Figure \ref{fig:dftb-aims} illustrates
the relative adsorption energies taking the C$_8$X$_2$ configuration as
reference for each doping level. As can be seen, the prediction of DFTB about
the doping level, at which the C$_8$X$_2$ configuration gets less stable as the
other investigated adsorption patterns is reliable. Furthermore, the DFTB prediction
about the change in the stability order for the sublattice polarized and
sublattice symmetric configurations is reliable as well.

\newpage

\bibliography{HaG}

\end{document}